\documentclass[10pt,a4paper]{article}
\usepackage{amsfonts,amssymb}
\usepackage[dvips]{graphicx}
\DeclareGraphicsExtensions{.eps}
\usepackage{epsfig}
\title
{Effects of Compton Cooling on the Hydrodynamic and 
the Spectral Properties of a Two Component Accretion Flow around a Black Hole}
\author
{Himadri Ghosh\thanks{himadri@bose.res.in}$^{1}$, Sudip K. Garain\thanks{sudip@bose.res.in}$^{1}$, Kinsuk Giri\thanks{kinsuk@bose.res.in}$^{1}$, Sandip K. Chakrabarti\thanks{chakraba@bose.res.in}$^{1,2}$\\
$^{1}$S. N. Bose National Centre for Basic Sciences, Salt Lake,
              Kolkata 700098, India\\
$^{2}$Indian Centre for Space Physics, Chalantika 43, Garia Station Rd., 
	     Kolkata, 700084, India} 
\begin{document}

\date{}


\maketitle

\label{firstpage}

\begin{abstract}
We carry out a time dependent numerical simulation where both the hydrodynamics and 
the radiative transfer are coupled together. We consider a two-component accretion 
flow in which the Keplerian disk is immersed inside an accreting low angular momentum 
flow (halo) around a black hole. The injected soft photons from the Keplerian 
disk are reprocessed by the electrons in the halo. We show that in presence of 
an axisymmetric soft-photon source, the spherically symmetric Bondi flow losses its 
symmetry and becomes axisymmetric. The low angular momentum flow was observed to slow 
down close to the axis and formed a centrifugal barrier which added new features into the 
spectrum. Using the Monte Carlo method, we generated the radiated spectra
as functions of the accretion rates. We find that the transitions from a hard state 
to a soft state is determined by the mass accretion rates of the disk and the halo. 
We separate out the signature of the bulk motion Comptonization and discuss its significance. 
We study how the net spectrum is contributed by photons suffering different number 
of scatterings and spending different amounts of time inside the Compton cloud.
We study the directional dependence of the emitted spectrum as well.
\end{abstract}



\section{Introduction}

The spectral and timing properties of a black hole candidate give away the most vital clues to the understanding the
nature of the invisible central object. The spectrum of radiation, particularly in high energies 
give information about the thermodynamic properties of matter accreting onto a black hole. The timing 
properties give information about how these thermodynamic properties are changing with time.
The thermodynamic properties such as the mass density, temperature etc. and the dynamic 
properties such as the velocity components are the solutions of the governing equations. 
Thus, a thorough knowledge of the spectral and timing properties are essential (e.g., Chakrabarti, 1996). 

There are several papers in the literature which have devoted themselves to study the 
spectral and timing properties of the accretion flows around black holes. Sunyaev \& Titarchuk (1980) suggested 
that the explanation of the emitted spectrum requires the presence of a Comptonizing hot electron plasma along with the
standard disk of Shakura \& Sunyaev (1973). There are several models in the literature,
such as the hot corona on a Keplerian disk (Haardt \& Maraschi, 1993),
unstable inner edge of the standard disk (Kobayashi et al. 2003),
hybrid EQPAIR model (Coppi, 1992) which uses both the thermal and non-thermal
electrons which empirically describe the nature of the possible Compton cloud. Other models
include those of Wandel and Liang (1991); Janiuk \& Czerny, 2000; Merloni \& Fabian, 2001; Zdziarski et al. 2003).
In the so-called Two Component Advective Flow (TCAF) model of Chakrabarti \& Titarchuk (1995), which is based on shock 
solutions in a sub-Keplerian flow (Chakrabarti, 1989), it was shown that 
the spectral properties are direct consequences of variation of accretion rates of the 
Keplerian (disk) and sub-Keplerian (halo) components. Subsequently, efforts were made to explain 
the timing properties. An important step in this direction is the theoretical work of Titarchuk and Lyubarskii (1995)
and Lyubarskii (1997) who showed the influence of noise and turbulences on the power density spectrum. Meanwhile, 
almost at the same time, Molteni, Sponholz and Chakrabarti (1996) pointed out that the resonance effects
between the cooling time scale and the infall time scale cause the Chakrabarti shocks (C-shocks)
to oscillate and cause the most important feature of the power density spectrum, namely, the 
quasi-periodic oscillations (QPOs). Molteni, Toth and Kuznetsov (1999) showed that these C-shocks
are actually stable even when azimuthal perturbations are given, though a vortex was shown to 
rotate anchoring the shocks, causing further enhancements in QPO power densities.
This was further expanded by Chakrabarti, Acharyya and Molteni (2004)
who relaxed the constraints on the equatorial symmetry and found that these 
shocks are prone to both vertical and radial oscillations of similar frequencies. Thus is it generally
established that the sub-Keplerian flows are responsible for both the spectral and timing properties
of the black hole candidates. This has been corroborated by several observations (Smith, Heindl 
\& Swank, 2002; Wu et al. 2002, Soria et al. 2001, Pottschmidt et al. 2006, Datta \& Chakrabarti, 2010). 

Given that the two component flows have been found to be useful to understand the spectral and
timing properties, it will be important to carry out the numerical simulations of radiative flows around black holes
which also include C-shocks. So far, however, only bremsstrahlung  or pseudo-Compton cooling have been added
into the time-dependent flows (Molteni et al. 1996; Chakrabarti et al. 2004, Proga 2007, Proga et al. 2008). 
Inclusion of the full-fledged Comptonization is prohibitively complex since 
the Comptonization efficiency depends on temperature and optical depth  
of the surrounding flow and this would depend on directions and time as well.  In the present paper,
we make the  first attempt to incorporate the time dependent simulation result which includes both 
hydrodynamics and radiative transfer. We use the low angular halo along with a Keplerian disk. We find how the 
Comptonization affects the temperature distribution of the flow and how this in turn affects the dynamics of 
the flow as well. So far, our solutions have been steady. We obtain the outgoing spectrum of radiation as well. 

In the next Section, we discuss the geometry of the soft photon source and the Compton cloud in our Monte Carlo 
simulations. The variation of the thermodynamic quantities and other
vital parameters are obtained inside the Keplerian disk and the Compton cloud which 
are required for the Monte Carlo simulations. In \S 3, we describe 
the simulation procedure and in \S 4, we present the
results of our simulations. Finally, in \S 5, we make concluding remarks.

\section{Geometry of the electron cloud and the soft photon source}

\begin{figure}
\centering{
\includegraphics[height=7truecm,width=7truecm,angle=0]{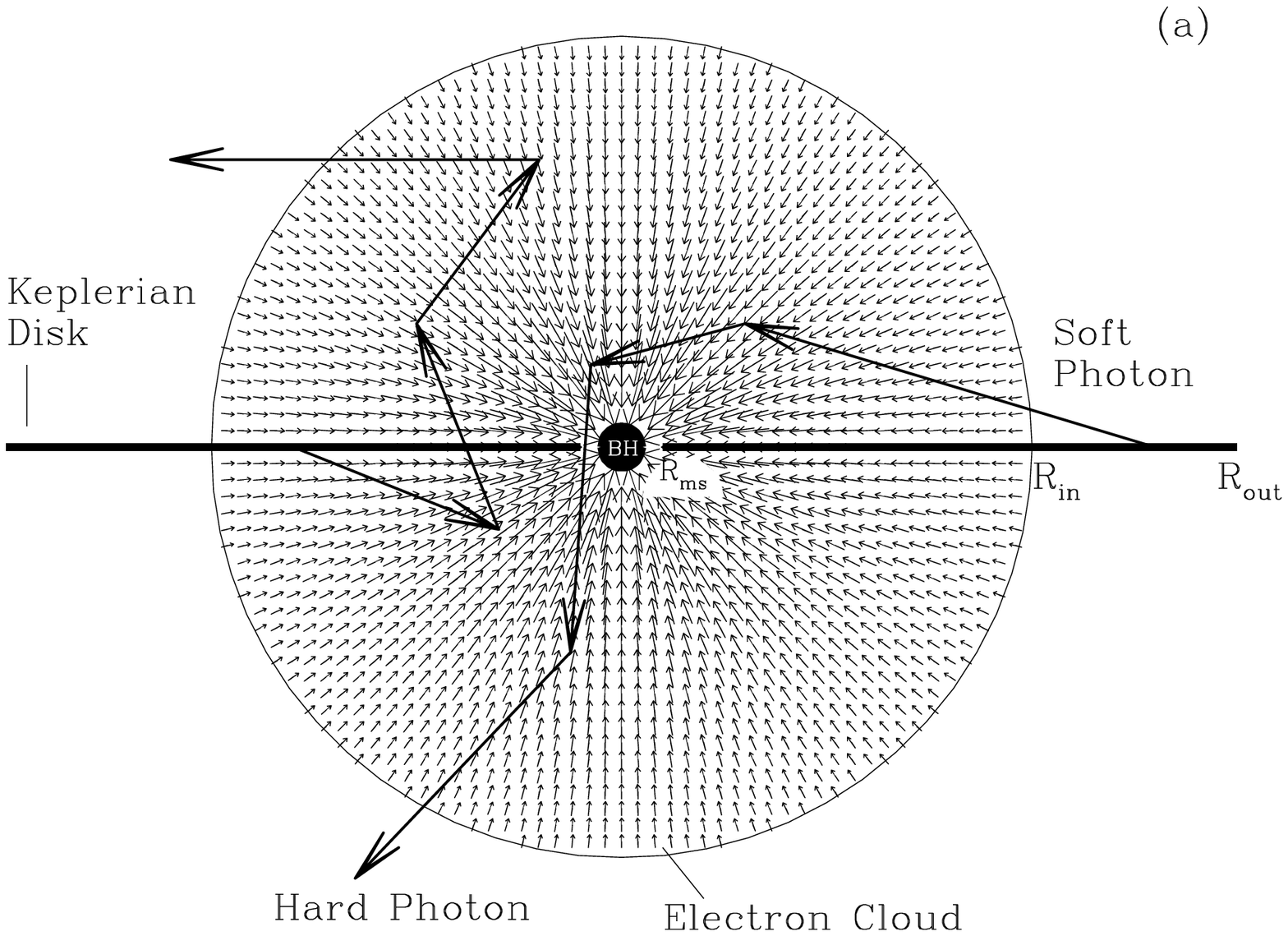}
\includegraphics[height=7truecm,width=7truecm,angle=0]{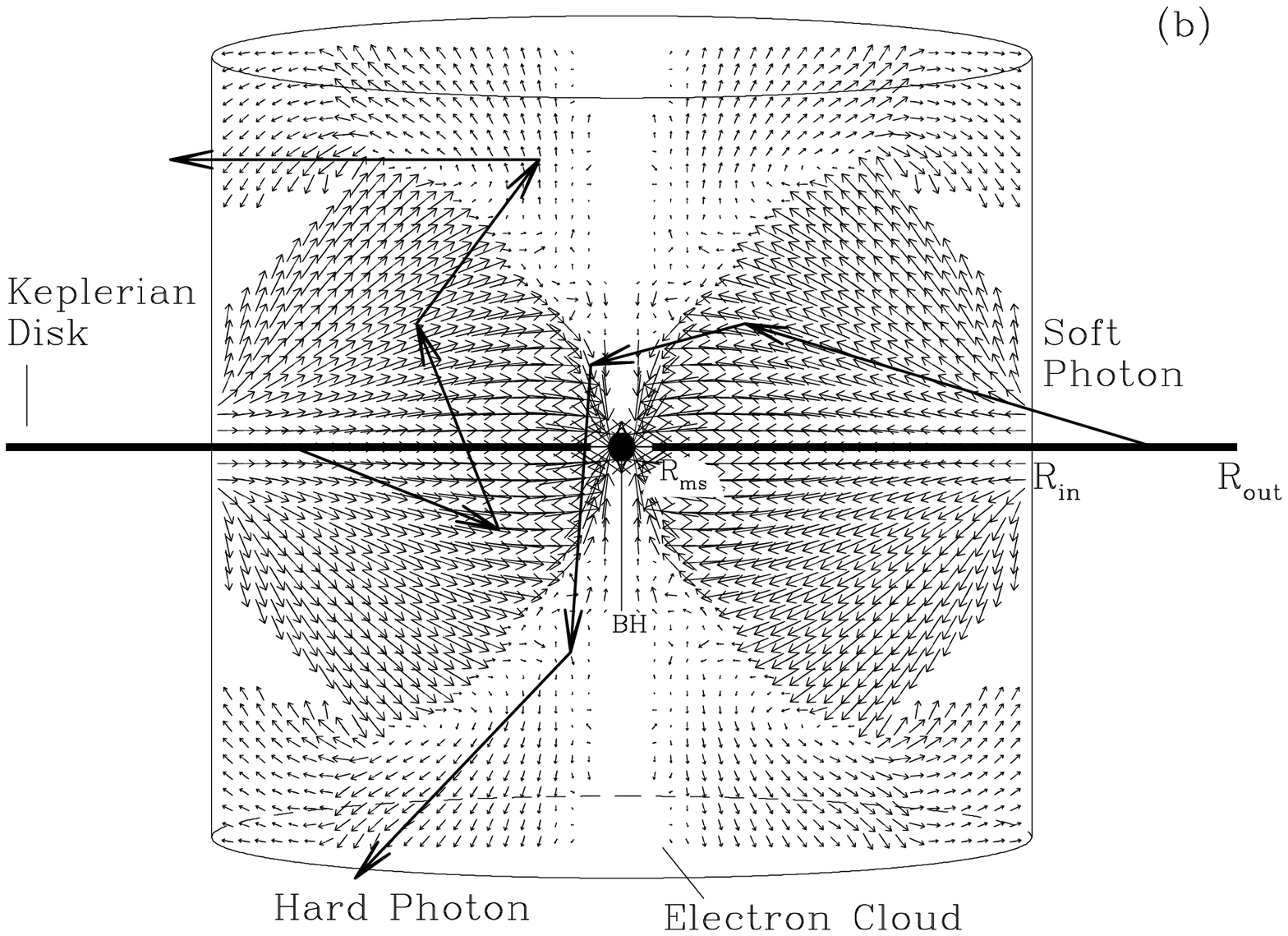}}
\caption{Schematic diagram of the geometry of our Monte Carlo simulations for (a) $\lambda=0$ 
and for (b) $\lambda=1$. Zigzag trajectories and velocity vectors are typical paths followed by 
the photons and the velocity vectors of the infalling matter inside the cloud.}
\label{fig1ab}
\end{figure}

In Figs. \ref{fig1ab}(a-b), we present cartoon diagrams of our simulation set up for 
(a) spherical Compton cloud (halo) with zero angular momentum (specific angular momentum  
i.e., angular momentum per unit mass, $\lambda=0$) and (b) rotating Compton cloud 
(halo) with a specific angular momentum $\lambda=1$. In the first case (a), we have the 
electron cloud within a sphere of radius $R_{in} = 200 r_g$, the Keplerian disk resides at the equatorial 
plane. The outer edge of this disk is located at $R_{out} = 300 r_g$ 
and it extends up to the marginally stable orbit $R_{ms} = 3 r_g$. 
At the centre of the sphere, a black hole of mass $10 M_{\odot}$ is located. The spherical matter 
is injected into the sphere from the radius $R_{in}$. It intercepts the soft 
photons emerging out of the Keplerian disk and reprocesses them via Compton or inverse    
Compton scattering. An injected photon may undergo a single, multiple or
no scattering at all with the hot electrons in between its emergence from the
Keplerian disk and its escape from the halo. 
The photons which enter the black holes are absorbed. In the second case (b), due to the 
presence of the angular momentum of the flow, the spherical symmetry of the flow is lost. 
The other parameters of the Keplerian disk and the halo remains the same as in (a).

\subsection{Distribution of temperature and density inside the Compton cloud}

\begin{figure}
\centering{
\includegraphics[height=6truecm,width=12truecm,angle=0]{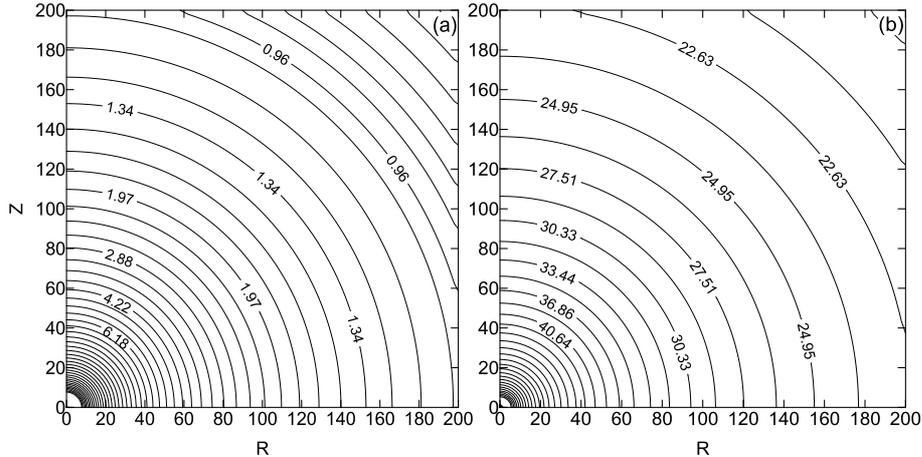}}
\caption{Density (a) and temperature (b) contours inside the spherical
halo in the absence of Compton cooling. Here, densities are in normalized 
unit and temperatures are in keV. $\lambda = 0$ is chosen. See text for details. }  
\label{fig2ab}
\end{figure}

A realistic accretion disk is expected to be three-dimensional. Assuming axisymmetry, 
we have calculated the flow dynamics using a finite difference method which uses the principle of 
Total Variation Diminishing (TVD) to carry out hydrodynamic simulations (see, Ryu, Chakrabarti \& Molteni, 1997
and references therein; Giri et al. 2010). At each time step, we carry out Monte Carlo simulation
to obtain the cooling/heating due to Comptonization. We incorporate the cooling/heating of 
each grid while executing the next time step of hydrodynamic simulation.
The numerical calculation for the two-dimensional flow has been carried out with 
$900 \times 900$ cells in a $200 r_g \times 200 r_g$ box. We chose the units in a way that
the outer boundary ($R_{in}$) is chosen to be unity and the matter density is normalized to become unity. 
We assume the black hole to be non-rotating and we use the pseudo-Newtonian potential $-\frac{1}{2(r-1)}$ 
(Paczy\'nski \& Wiita, 1980) to calculate the flow geometry around a black hole 
(Here, $r$ is in the unit of Schwarzschild radius $r_g=2GM_{bh}/c^2$). Velocities and angular 
momenta are measured in units of $c$, the velocity of light and $r_g c$ respectively.
In Figs. \ref{fig2ab}(a-b) we show the snapshots of the density 
and temperature (in keV) profiles obtained in a steady state purely from our hydrodynamic
simulation. The density contour levels are drawn for $0.65 - 1.01$ (levels increasing 
by a factor of $1.05$) and $1.01 - 66.93$ (successive level
ratio is $1.1$). The temperature contour levels are drawn for $16.88 - 107.8$ keV
(successive level ratio is $1.05$).

\subsection{Properties of the Keplerian disk}

The soft photons are produced from a Keplerian disk whose inner edge has been kept 
fixed at the marginally stable orbit $R_{ms}$, while the outer edge is located at $R_{out}$ (=assumed to 
be at $300 r_g$ in this paper). The source of the soft photons have a multicolor blackbody 
spectrum coming from a standard (Shakura \& Sunyaev, 1973, hereafter SS73) disk. We assume the disk 
to be optically thick and the opacity due to free-free absorption is more 
important than the opacity due to scattering. The emission is black body type 
with the local surface temperature (SS73):
\begin{equation} 
T(r) \approx 5 \times 10^7 (M_{bh})^{-1/2}(\dot{M_d}_{17})^{1/4} (2r)^{-3/4}
 \left[1- \sqrt{\frac{3}{r}}\right]^{1/4} K ,
\label{eq:a1}
\end{equation}
The total number of photons emitted from the disk surface is obtained by integrating
over all frequencies ($\nu$) and is given by,
\begin{equation}
n_\gamma(r) = \left[16 \pi \left( \frac{k_b}{h c} \right)^3 \times 1.202057 \right]
\left(T(r)\right)^3
\label{eq:a2}
\end{equation}
The disk between radius $r$ to $r+\delta r$ injects $dN(r)$ number of soft photons.
\begin{equation}
dN(r) =  4 \pi r \delta r H(r) n_\gamma(r),
\label{eq:a3}
\end{equation}
where, $H(r)$ is the half height of the disk given by:
\begin{equation} 
H(r) = 10^5 \dot{M_d}_{17} \left[1- \sqrt{\frac{3}{r}}\right] \rm{cm}.
\label{eq:a4}
\end{equation}
In the Monte Carlo simulation, we incorporated the directional effects of photons 
coming out of the Keplerian disk with the maximum number of photons emitted in the 
$z$-direction and minimum number of photons are generated along the plane of the disk. 
Thus, in the absence of photon bending effects, the disk is invisible as seen 
edge on. The position of each emerging photon is   
randomized using the distribution function (Eq. 3). In the above equations, the mass 
of the black hole $M_{bh}$ is measured in units of the mass of the Sun ($M_\odot$), 
the disk accretion rate $\dot{M_d}_{17}$ is in units of $10^{17}$ gm/s. We chose 
$M_{bh} = 10$ in the rest of the paper.

\section{Simulation Procedure}

In a given run, we assume a Keplerian disk rate ($\dot{m}_d$) and a 
sub-Keplerian halo rate ($\dot{m}_h$). The specific energy ($\epsilon$) of 
the halo provides the hydrodynamic (e.g., number density of the
electrons and the velocity distribution) and the thermal properties of matter. Since we
chose the Paczy\'nski-Wiita (1980) potential, the radial velocity is not exactly 
unity at $r=1$, the horizon. It becomes unity just outside. In order not 
to over estimate the effects of bulk motion Comptonization (Chakrabarti \& Titarchuk, 1995) 
which is due to the momentum transfer of the moving electrons to the horizon, we kept the 
highest velocity to be 1. We use the absorbing boundary condition at $r=1.5$ ($\lambda=0$ case) 
and $r=2.5$ ($\lambda=1$ case). These simplifying assumptions do not 
affect our conclusions, especially because we are studying inviscid flow and the specific
angular momentum is constant. Photons are generated from the Keplerian disk as mentioned 
before and may be intercepted by the sub-Keplerian halo (sphere in Fig.  
\ref{fig1ab}a and cylinder in Fig. \ref{fig1ab}b).  

To begin the Monte Carlo code, we randomly generated soft photons from the Keplerian disk.
The energy of the soft photon at radiation temperature $T(r)$ is calculated using the
Planck's distribution formula, where the number density of the photons
($n_\gamma(E)$) having an energy $E$ is expressed by,
\begin{equation}
n_\gamma(E) = \frac{1}{2 \zeta(3)} b^{3} E^{2}(e^{bE} -1 )^{-1}, 
\label{eq:a5}
\end{equation}
where, $b = 1/kT(r)$ and $\zeta(3) = \sum^\infty_1{l}^{-3} = 1.202$, the Riemann zeta function.

Using another set of random numbers we obtained the direction of the injected photon and with yet
another random number we obtained a target optical depth $\tau_c$ at which the scattering takes place. The photon was followed within the electron cloud till the optical depth ($\tau$) 
reached $\tau_c$. The increase in optical depth ($d\tau$) during its traveling  of
a path of length $dl$ inside the electron cloud is given by: $d\tau = \rho_n \sigma dl$, where
$\rho_n$ is the electron number density.

The total scattering cross section $\sigma$ is given by Klein-Nishina formula:
\begin{equation}
\sigma = \frac{2\pi r_{e}^{2}}{x}\left[ \left( 1 - \frac{4}{x} - \frac{8}{x^2} \right) ln\left( 1 + x \right) + \frac{1}{2} + \frac{8}{x} - \frac{1}{2\left( 1 + x \right)^2} \right],
\label{eq:a6}
\end{equation}
where, $x$ is given by,
\begin{equation}
x = \frac{2E}{m c^2} \gamma \left(1 - \mu \frac{v}{c} \right),
\label{eq:a7}
\end{equation}
$r_{e} = e^2/mc^2$ is the classical electron radius and $m$ is the mass of the electron.

We have assumed here that a photon of energy $E$ and momentum $\frac{E}{c}\bf{\widehat{\Omega}}$
is scattered by an electron of energy $\gamma mc^{2}$ and momentum $\overrightarrow{\bf{p}} = \gamma m \overrightarrow{\bf{v}}$, with $\gamma = \left( 1 - \frac{v^2}{c^2}\right)^{-1/2}$ and $\mu = \bf{\widehat{\Omega}}. \widehat{\bf{v}}$.
At this point, a scattering is allowed to take place. The photon selects an electron and the energy
exchange is computed using the Compton or inverse Compton scattering formula. The electrons
are assumed to obey relativistic Maxwell distribution inside the Compton cloud.
The number $dN(p)$ of Maxwellian electrons having momentum between
$\vec{p}$ to $\vec{p} + d\vec{p}$ is expressed by,
\begin{equation}
dN(\vec{p}) \propto exp[-(p^2c^2 + m^2c^4)^{1/2}/kT_e]d\vec{p}.
\label{eq:a8}
\end{equation}

We take a steady state flow profile from a hydrodynamics code to start the Monte Carlo simulation. 
When a photon interacts with an electron via Compton or inverse-Compton scattering, it 
loses or gains some energy ($\Delta E$). At each grid point, we compute $\Delta E$.
We update the energy of  the flow at this grid by this amount and continue the hydrodynamic code with 
this modified energy. This, in turn, modify the hydrodynamic profile. Thus the Monte Carlo 
code for radiative transport and numerical code are coupled together. In case the final 
state is steady, the temperature of the cloud would be reduced progressively 
to a steady value from the initial state where no cooling was assumed. 
If the final state is oscillatory, the solution would settle into a state with Comptonization.

\subsection{Details of the Coupling procedure}

Once a steady state is achieved in the non-radiative hydro-code, we compute the spectrum using the 
Monte Carlo code. This is the spectrum in the first approximation. To include cooling in the coupled 
code, we follow these steps: (a) we calculate the velocity, density and temperature profiles of the 
electron cloud from the output of the hydro-code. (b) Using the Monte Carlo code we calculate the spectrum. 
(c) Electrons are cooled (heated up) by the inverse-Compton (Compton) scattering. We calculate the 
amount of heat loss (gain) by the electrons and its new temperature and energy distributions and
(d) taking the new temperature and energy profiles as initial condition, we run the hydro-code for a period of time.
Subsequently, we repeat the steps (a-d). In this way, we get an opportunity to 
see how the spectrum is modified as the iterations proceed. The iterations
stop when two successive steps produce virtually the same temperature profile and the emitted spectrum.

\subsubsection{Calculation of energy reduction using Monte Carlo code:}

For Monte Carlo simulation, we divide the Keplerian disk in different annuli of width $D(r)=0.5$. 
Each annulus is characterized by its central temperature $T(r)$. The total number of photons emitted 
from the disk surface of each annulus can be calculated using Eqn. (\ref{eq:a3}).
This total number comes out to be $\sim~10^{39-40}$ for $\dot{m}_d = 1.0$.
In reality, one cannot inject this much number of photons in Monte Carlo simulation
because of the limitation of computation time. So we replace this large number of
by a low number of bundles, say, $N_{comp}(r)~\sim~10^7$ and calculate a weightage 
factor 
$$
f_W = \frac{dN(r)}{N_{comp}(r)}.
$$ 
Clearly, from each annulus, the number of photons in a bundle will vary. This is computed exactly and 
used to compute the change of energy due to Comptonization. When this injected photon is inverse-Comptonized 
(or, Comptonized) by an electron
in a volume element of size $dV$, we assume that $f_W$ number of photons has suffered similar
scattering with the electrons inside the volume element $dV$. If the energy
loss (gain) per electron in this scattering is $\Delta E$, we multiply this
amount by $f_W$ and distribute this loss (gain) among all the electrons inside 
that particular volume element. This is continued for all the $10^7$ bundles of photons
and the revised energy distribution is obtained.

\subsubsection{Computation of the temperature distribution after cooling} 

Since the hydrogen plasma considered here is ultra-relativistic ($\gamma=\frac{4}{3}$ throughout 
the hydrodynamic simulation), thermal energy per particle is $3k_BT$ where $k_B$ is 
Boltzmann constant, $T$ is the temperature of the particle.  
The electrons are cooled by the inverse-Comptonization of the soft photons
emitted from the Keplerian disk. The protons are cooled because of the 
Coulomb coupling with the electrons. Total number of electrons inside 
any box with the centre at location $(ir,iz)$ is given by,
\begin{equation}
dN_e(ir,iz) = 4\pi rn_e(ir,iz)drdz, 
\label{eq:a9}
\end{equation}
where, $n_e(ir,iz)$  is the electron number density at $(ir,iz)$ location, and
$dr$ and $dz$ represent the grid size along $r$ and $z$ directions respectively. So
the total thermal energy in any box is given by $3k_BT(ir,iz)dN_e(ir,iz) = 12\pi
rk_BT(ir,iz)n_e(ir,iz)drdz,$ where $T(ir,iz)$ is the temperature at $(ir,iz)$
grid. We calculate the total energy loss (gain) $\Delta E$ of electrons inside the
box according to what is presented above and subtract that amount to get the
new temperature of the electrons inside that box as 
\begin{equation}k_BT_{new}(ir,iz) = 
k_BT_{old}(ir,iz)-\frac{\Delta E}{3dN_e(ir,iz)}.\label{eq:a10} \end{equation}

\subsection{Details of the hydrodynamic simulation code}

As mentioned above, after every spell of cooling by the Monte Carlo code for a very short time
step, the hydro-code is run without assuming cooling. This procedure is repeated. While running the
hydro-code the following process is followed.

To model the initial injection of matter, we consider an axisymmetric flow of gas in the pseudo-Newtonian 
gravitational field of a black hole of mass $M_{bh}$ located at the centre in the cylindrical coordinates  
$[R,\theta,z]$. We assume that at infinity, the gas pressure is negligible and the
energy per unit mass vanishes. We also assume that the gravitational field 
of the black hole can be described by Paczy\'{n}ski \& Wiita (1980),
$$
\phi(r) = -{GM_{bh}\over(r-r_g)}, 
$$
where, $r=\sqrt{R^2+z^2}$, and the Schwarzschild radius is given by, 
$$
r_g=2GM_{bh}/c^2 .
$$
We also assume a polytropic equation of state for the
accreting (or, outflowing) matter, $P=K \rho^{\gamma}$, where,
$P$ and $\rho$ are the isotropic pressure and the matter density
respectively, $\gamma$ is the adiabatic index (assumed 
to be constant throughout the flow, and is related to the
polytropic index $n$ by $\gamma = 1 + 1/n$) and $K$ is related
to the specific entropy of the flow $s$. The details of the code is 
described in Ryu, Molteni \& Chakrabarti (1997) and in Giri et al. (2010). 

Our computational box occupies one quadrant of the R-z plane with $0 \leq R \leq 200$ and $0 \leq z \leq 200$. 
The incoming gas enters the box through the outer boundary, located at $R_{in} = 200$. We have chosen the density 
of the incoming gas ${\rho}_{in} = 1$ for convenience since, in the absence of self-gravity
and cooling, the density is scaled out, rendering the simulation results valid for any accretion rate.
As we are considering only energy flows while keeping the boundary of the numerical grid at a finite distance, 
we need the  sound speed $a$ (i.e., temperature) of the flow and the incoming velocity at the boundary points. 
For the spherical flow with zero angular momentum (Bondi flow), we have taken the boundary values from standard
pseudo-Bondi solution. We injected the matter from both the outer boundary  of R and z coordinate.
In order to mimic the horizon of the black hole at the Schwarzschild radius, we placed an absorbing inner boundary
at $r = 1.5 r_g$, inside which all material is completely absorbed into the black hole. For the background matter
(required to avoid division by zero)
we used a stationary gas with density ${\rho}_{bg} = 10^{-6}$ and sound speed (or temperature) the same as that of the incoming
gas. Hence the incoming matter has a pressure $10^6$ times larger than that of the background matter. All
the calculations were performed with $900 \times 900$ cells, so each grid has a size of $0.22$ 
in units of the Schwarzschild radius.

All the simulations are carried out assuming a stellar mass black hole $(M = 10{M_\odot})$. 
The procedures remain equally valid for massive/super-massive black holes.
We carry out the simulations till several thousands of dynamical time-scales are passed. 
In reality, this corresponds to a few seconds in physical units.

\section{Results and Discussions}

In Table 1, we summarize all the cases for which the simulations have been presented 
in this paper. In Column 1, various cases are marked. Columns 2 and 3 give the angular momentum 
($\lambda$) and the specific energy ($\epsilon$) of the flow. The Keplerian disk rate ($\dot{m}_d$)
and the sub-Keplerian halo rate ($\dot{m}_h$) are listed in Columns 4 and 5. The number of soft photons, 
injected from the Keplerian disk ($N_{inj}$) for various disk rates can be found in Column 6. 
Column 7 lists the number of photons ($N_{sc}$) that have suffered at least one scattering 
inside the electron cloud. The number of photons ($N_{unsc}$), escaped from the cloud without 
any scattering are listed in Column 8. Columns 9 and 10 give the percentages of injected photons that 
have entered into the black hole ($N_{bh}$) and suffered scattering ($p = \frac{N_{sc}}{N_{inj}}$), 
respectively. The cooling time ($t_0$) of the system is defined as the expected time for the 
system to lose all its thermal energy with the particular flow parameters (namely, $\dot{m}_d$ and $\dot{m}_h$).
We calculate $t_0 = E/{\dot{E}}$ in each time step, where, $E$ is the total energy content of the system and 
${\dot{E}}$ is the energy gain or loss by the system in that particular time step. We present the energy 
spectral index $\alpha$ $\left[I(E) \sim E^{-\alpha}\right]$ obtained from our simulations in the last column.

{\small{
\begin {tabular}[h]{|c|c|c|c|c|c|c|c|c|c|c|c|}
\multicolumn{12}{|c|}{Table 1: Parameters used for the simulations and a summary of results.}\\
\hline Case & $\lambda$ & $\epsilon$ & $\dot{m}_d$ & $\dot{m}_h$ & $N_{inj}$ & 
$N_{sc}$ & $N_{unsc}$ & $N_{bh}$ [$\%$] & $p$ [$\%$] & $t_0$ [sec] & $\alpha$\\
\hline
1a & 0 & 22E-4 & 1   & 1   & 4.3E+40 & 8.7E+39 & 3.5E+40 & 0.119 & 20.030 & 228.3 & 1.15, 0.99 \\
1b & 0 & 22E-4 & 2   & 1   & 1.5E+41 & 2.9E+40 & 1.2E+41 & 0.120 & 20.023 & 63.6  & 1.30, 1.0  \\
1c & 0 & 22E-4 & 5   & 1   & 7.3E+41 & 1.5E+41 & 5.9E+41 & 0.121 & 19.942 & 12.4  & 1.40, 0.96 \\
1d & 0 & 22E-4 & 10  & 1   & 2.5E+42 & 5.0E+41 & 2.0E+42 & 0.121 & 19.816 & 4.2   & 1.65, 0.90 \\
1e & 0 & 22E-4 & 1   & 0.5 & 4.3E+40 & 4.7E+39 & 3.9E+40 & 0.070 & 10.886 & 380.0 & 1.57       \\
1f & 0 & 22E-4 & 1   & 2   & 4.3E+40 & 1.5E+40 & 2.8E+40 & 0.230 & 34.324 & 118.9 & 1.1        \\
1g & 0 & 22E-4 & 1   & 5   & 4.3E+40 & 2.6E+40 & 1.8E+40 & 0.502 & 59.012 & 48.0  & 0.7        \\
1h & 0 & 22E-4 & 1   & 10  & 4.3E+40 & 3.3E+40 & 1.1E+40 & 0.699 & 75.523 & 35.1  & 0.45       \\
\hline
2a & 1 & 3E-4 & 1  & 1    & 6.3E+40 & 1.2E+40 & 5.1E+40 & 0.285 & 19.199 & 79.7 & 0.88 \\
2b & 1 & 3E-4 & 2  & 1    & 2.1E+41 & 4.1E+40 & 1.7E+41 & 0.283 & 19.278 & 21.9 & 0.94 \\
2c & 1 & 3E-4 & 5  & 1    & 1.0E+42 & 1.9E+41 & 8.1E+41 & 0.283 & 19.205 & 4.3  & 1.03 \\
2d & 1 & 3E-4 & 10 & 1    & 3.6E+42 & 6.9E+41 & 2.9E+42 & 0.289 & 18.941 & 1.4  & 1.17 \\
2e & 1 & 3E-4 & 10 & 0.5  & 3.6E+42 & 3.9E+41 & 3.2E+42 & 0.190 & 10.768 & 1.9  & 1.37 \\
2f & 1 & 3E-4 & 10 & 1.5  & 3.6E+42 & 9.3E+41 & 2.7E+42 & 0.372 & 25.492 & 1.1  & 1.01 \\
2g & 1 & 3E-4 & 10 & 2    & 3.6E+42 & 1.1E+42 & 2.5E+42 & 0.443 & 30.758 & 0.9  & 0.95 \\
2h & 1 & 3E-4 & 10 & 5    & 3.6E+42 & 1.8E+42 & 1.7E+42 & 0.690 & 51.073 & 0.7  & 0.59 \\
\hline
\end{tabular}
}}

\subsection{Compton cloud with no angular momentum}

\begin{figure}
\centering{
\includegraphics[height=12truecm,width=12truecm,angle=0]{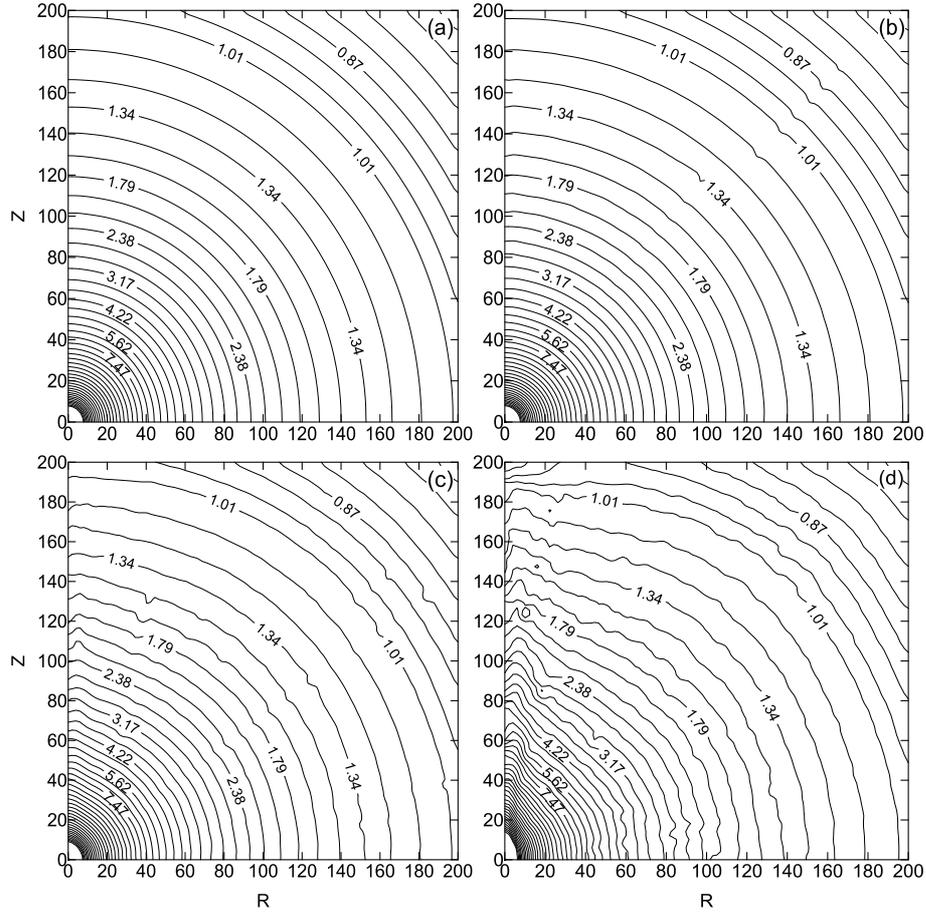}}
\caption{Changes in the density distribution in presence of cooling. $\lambda = 0$ and 
$\dot{m}_h = 1$ for all the cases. Disk accretion rate $\dot{m}_d$ used 
are (a) 1, (b) 2, (c) 5 and (d) 10 respectively (Cases 1(a-d) of Table 1). 
The density contours are drawn using the same contour levels as in Fig. \ref{fig2ab}a.}
\label{fig3abcd}
\end{figure}

\begin{figure}
\centering{
\includegraphics[height=12truecm,width=12truecm,angle=0]{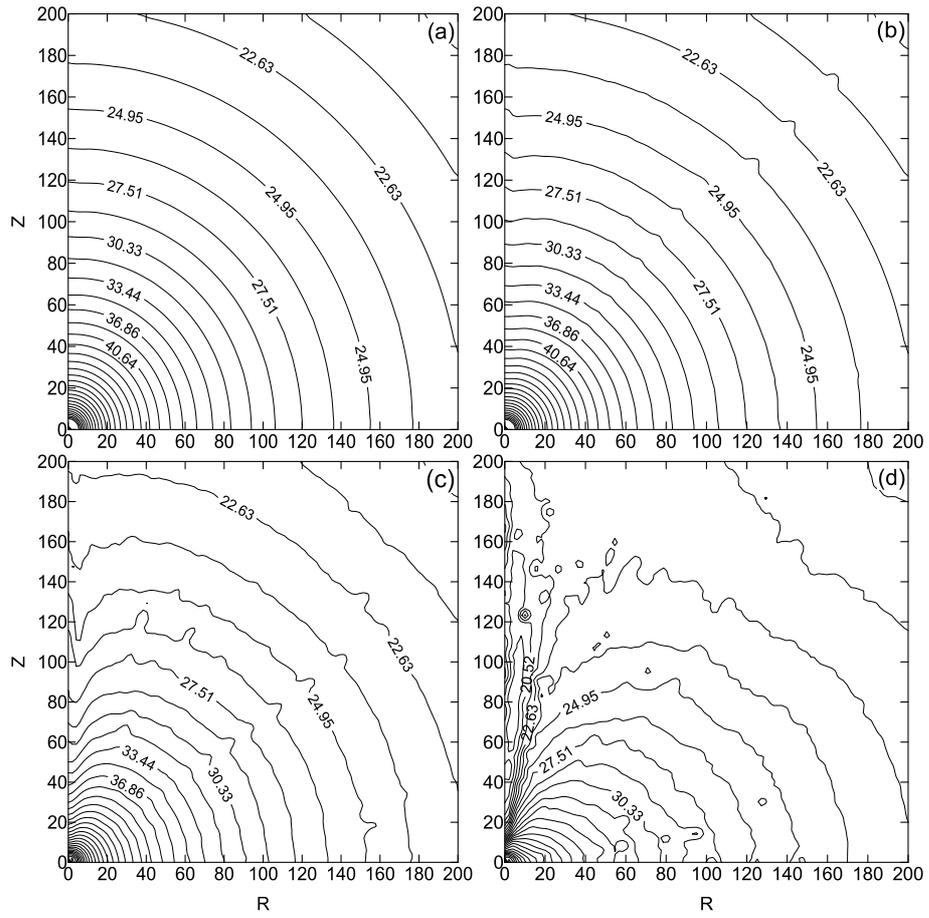}}
\caption{Changes in the temperature distribution in presence of cooling. 
$\lambda = 0$ and $\dot{m}_h = 1$ for all the cases. Disk 
accretion rate $\dot{m}_d$ is (a)1, (b)2, (c)5 and (d)10 
respectively (Cases 1(a-d) of Table 1). Contours are drawn using the same levels as in Fig. \ref{fig2ab}b.}
\label{fig4abcd}
\end{figure}

First we discuss the results corresponding to the Cases 1(a-d) of Table 1. 
In Fig. \ref{fig3abcd}(a-d) we present the changes in density distribution 
as the disk accretion rates are changed: $\dot{m}_d =$ (a) 1, (b) 2, 
(c) 5 and (d) 10 respectively. We notice that as the accretion rate 
of the disk is enhanced, the density distribution losses its spherical symmetry. In particular the 
density at a given radius is enhanced in a conical region region along the axis. This is due to 
the cooling of the matter by  Compton scattering. To show this, in Fig. \ref{fig4abcd}(a-d) 
we show the contours of constant temperatures (marked on curves) of the same four cases. 
We notice that the temperature is reduced along the axis (where the optical depth as seen by the 
soft photons from the Keplerian disk is higher) drastically after repeated Compton scattering. 

\begin{figure}
\vspace {0.4cm}
\centering{
\includegraphics[height=5.7truecm,width=5.7truecm,angle=0]{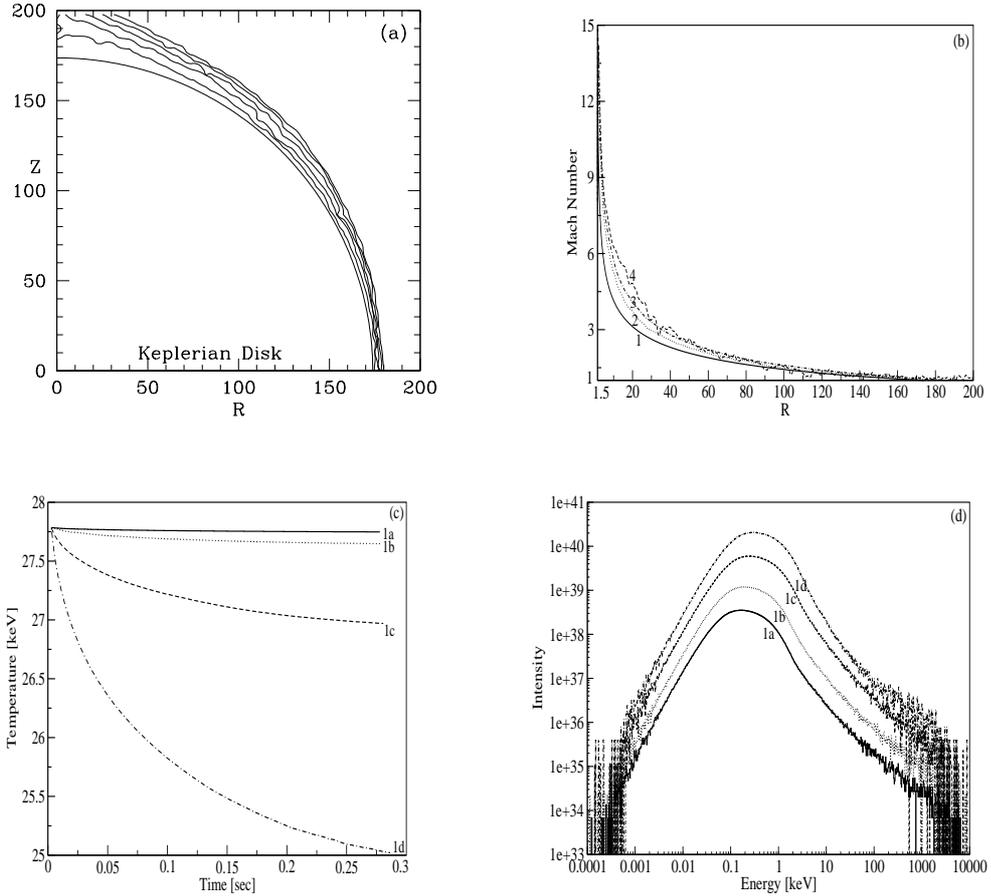}
\hskip 1.5cm
\includegraphics[height=5.3truecm,width=5.5truecm,angle=0]{Fig5b.eps}}\\
\vspace {1.0cm}
\centering{
\includegraphics[height=5.3truecm,width=5.3truecm,angle=0]{Fig5c.eps}
\hskip 1.5cm
\includegraphics[height=5.3truecm,width=6.0truecm,angle=0]{Fig5d.eps}}
\caption{(a) Sonic surfaces at different stages of iterations. The final curve
represents the converged solution. The initial spherical sonic surface become 
prolate spheroid due to cooling by the Keplerian disk at the equatorial plane. 
Parameters are for Case 1d (Table 1).
(b) Mach number variation as a function of distance after a complete solution of 
the radiative flow is obtained. Plot no. 1 corresponds to the solution from adiabatic Bondi flow. 
Plots 2-4 are the solutions along the equatorial plane, the diagonal and the axis of the disk. 
Parameters are for Case 1d (Table 1). 
(c) Variation of the average temperature of the Compton cloud as the iteration proceeds when 
the disk accretion rate is varied. $\dot{m}_h = 1$. The solid, dotted, dashed and dash-dotted plots are 
for $\dot{m}_d = 1$, $2$, $5$ and $10$ respectively. Case numbers (Table 1) are marked. 
With the increase of disk rate, the temperature of the Compton cloud converges to a lower temperature.
(d) Variation of the spectrum with the increase of disk accretion rate. Parameters are the same as in (c).
With the increase in $\dot{m}_d$, the intensity of the spectrum increases due to the increase 
in $N_{inj}$ (see, Table 1). The spectrum is softer for the higher value of $\dot{m}_d$. 
Spectral slopes for each of these spectra are listed in Table 1.}
\label{fig5abcd}
\end{figure}

In Fig. \ref{fig5abcd}(a-d), we show the hydrodynamic and radiative properties. 
In Fig. \ref{fig5abcd}(a), we show the sonic surfaces. The lowermost curve corresponds 
to theoretical solution for an adiabatic flow (e.g., Chakrabarti, 1990). Other 
curves from the bottom to the top are the iterative solutions for the Case 1d mentioned above. 
As the disk rate is increased, the cooling increases, and consequently, the Mach number 
increases along the axis. Of course, there are other effects: The cooling causes the 
density to go up to remain in pressure equilibrium. In Fig. \ref{fig5abcd}b, 
the Mach number variation is shown. The lowermost curve (marked 1) is from the theoretical 
consideration. Plots 2-4 are the variation of Mach number with radial distance 
along the equatorial plane, along the diagonal and along the vertical axis 
respectively. In Fig. \ref{fig5abcd}c, the average temperature of the spherical halo is plotted as a function of the
iteration time until almost steady state is reached. The cases are marked on the curves. We note that as the 
injection of soft photons increases, the average temperature of the halo decreases drastically. 
In Fig. \ref{fig5abcd}d, we have plotted the energy dependence of the photon intensity.
We find that, as we increase the disk rate, keeping the halo rate fixed, number 
of photons coming out of the cloud in a particular energy bin increases and the spectrum becomes softer. 
This is also clear from Table 1, $N_{inj}$ increases with $\dot{m}_d$, increasing $\alpha$. We find the 
signature of double slope in these cases. As the disk rate increases, the second slope becomes steeper. 
This second slope is the signature of bulk motion Comptonization. As $\dot{m}_d$ increases, the cloud becomes 
cooler (Plot 5c) and the power-law tail due to the bulk motion Comptonization (Chakrabarti \& Titarchuk, 1995) becomes prominent. 

\begin{figure}
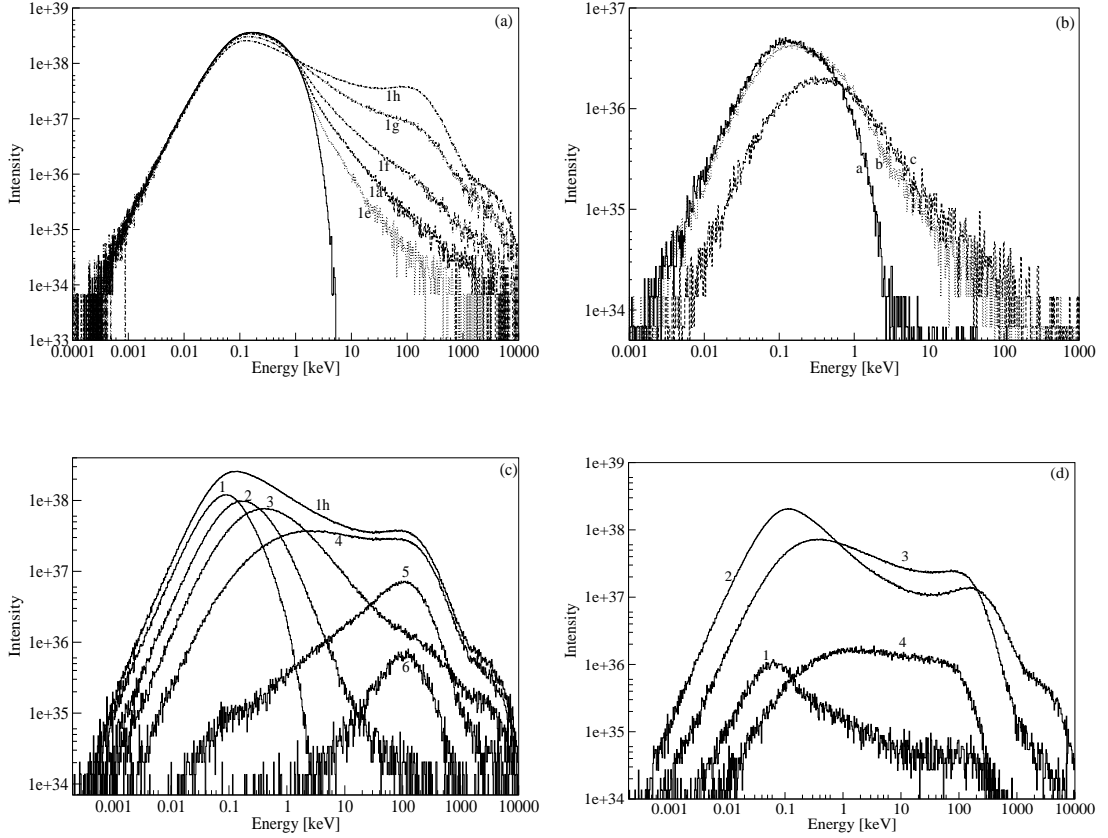

\vspace {0.4cm}
\centering{
\includegraphics[height=5truecm,width=7truecm,angle=0]{Fig6a.eps}
\hskip 0.2cm
\includegraphics[height=5truecm,width=7truecm,angle=0]{Fig6b.eps}}\\
\vspace {1.0cm}
\centering{
\includegraphics[height=5truecm,width=7truecm,angle=0]{Fig6c.eps}
\hskip 0.2cm
\includegraphics[height=5truecm,width=7truecm,angle=0]{Fig6d.eps}}
\caption{(a) Variation of the spectrum with the increase of the halo accretion rate, keeping the 
disk rate ($\dot{m}_d = 1$) and angular momentum of the flow ($\lambda = 0$) fixed. 
The dotted, dashed, dash-dotted, double dot-dashed and double dash-dotted curves show the 
spectra for $\dot{m}_h = 0.5$, $1$, $2$, $5$ and $10$ respectively. 
The injected multicolor blackbody spectrum supplied by the Keplerian disk is shown (solid line). 
(b) Directional dependence of the spectrum: $\lambda = 0$, $\dot{m}_h = 2$, $\dot{m}_d = 1$  are the
flow parameters. The solid, dotted and dashed curves are for observing angles $2^{\circ}$, $45^{\circ}$ 
and $90^{\circ}$ respectively. All the angles are measured with respect to the rotation axis ($z$-axis). 
Intensity of spectra emerging from the cloud after suffering various number of scatterings (c) and 
 at four different times (d) immediately after the injection of soft photons. Case 1h is assumed. 
The spectra of the photons suffering 0, 1-2, 3-6, 7-18, 19-28 and more than 29 scatterings are 
shown by the plots 1, 2, 3, 4, 5 and 6 (Fig. \ref{fig6abcd}c) respectively, within the cloud. Curve 1h is the net 
spectrum for which these components are drawn. As the number of scattering increases, the 
photons gain more and more energy from the hot electron cloud through inverse Comptonization 
process. The spectra of the photons spending 0.01-1, 1-40, 40-100 and more than 100 ms time 
inside the electron cloud are marked by 1, 2, 3 and 4 (Fig. \ref{fig6abcd}d) respectively.}
\label{fig6abcd}
\end{figure}

In Fig. \ref{fig6abcd}a we show the variation of the energy spectrum with the increase of the halo 
accretion rate, keeping the disk rate ($\dot{m}_d = 1$) and angular momentum of the flow ($\lambda = 0$) fixed. 
The injected multi-color blackbody spectrum supplied by the Keplerian disk is shown (solid line).
The dotted, dashed, dash-dotted, double dot-dashed and double dash-dotted curves show the 
spectra for $\dot{m}_h = 0.5$, $1$, $2$, $5$ and $10$ respectively. 
The injected multicolor blackbody spectrum supplied by the Keplerian disk is shown (solid line). 
The spectrum becomes harder for higher values of $\dot{m}_h=1$ as it is difficult to cool a higher density 
matter with the same number of injected soft photon. In Fig. \ref{fig6abcd}b, we show the 
directional dependence of the spectrum. for $\lambda = 0$, $\dot{m}_h = 2$, 
$\dot{m}_d = 1$ (Case 1f). The solid, dotted and dashed curves are for 
observing angles (a) $2^{\circ}$,  (b) $45^{\circ}$ and (c) $90^{\circ}$ respectively. 
All the angles are measured with respect to the rotation axis ($z$-axis). 
As expected, the photons arriving along the z-axis would be dominated by the soft photons from the 
Keplerian disk while the power-law would dominate the spectrum coming edge-on.

We now study the dependence of spectrum on the time delay between the injected photon and the outgoing photon. 
Depending on the number of scatterings suffered and the length of path traveled, different 
photons spend different times inside the Compton cloud. The energy gain or loss by any 
photon depends on this time. Fig. \ref{fig6abcd}c shows the spectrum of the photons suffering 
different number of scatterings inside the cloud. Here 1, 2, 3, 4, 5 and 6 shows 
the spectrum for 6 different ranges of number of scatterings. Plot 1 shows the spectrum of the 
photons that have escaped from the cloud without suffering any scattering. This spectrum is 
nearly the same as the injected spectrum, only difference is that it is Doppler shifted. As the 
number of scattering increases (spectrum 2, 3 and 4), the photons are more and more energized
via inverse Compton scattering with the hot electron cloud. For 
scatterings more than 19, the high energy photons start loosing energy through Compton scattering 
with the relatively lower energy electrons. Components 5 and 6 show the spectra of the 
photons suffering 19-28 scatterings and the photons 
suffering more than 28 respectively. Here the flow parameters are: 
$\dot{m}_d = 1$, $\dot{m}_h = 10$ and $\lambda = 0$ (Case 1h Table 1).

In Fig. \ref{fig6abcd}d, we plot the spectrum emerging out of the electron cloud at four different time ranges. 
In the simulation, that the photons take $0.01$ to $130$ ms to come out of the system. We divide this time range 
into 4 suitable bins and plot their spectrum. Case 1h of Table 1 is considered. We observe that the spectral 
slopes and intensities of the four spectra are different. As the photons spend more and more time inside the 
cloud, the spectrum gets harder (plots 1, 2 and 3). However, very high energy photons which spend maximum 
time inside the cloud lose some energy to the relatively cooler electrons before escaping from the cloud. Thus 
the spectrum 4 is actually the spectrum of Comptonized photons.

\subsection{Compton cloud  with very low angular momentum}

\begin{figure}
\centering{
\includegraphics[height=8truecm,width=15truecm,angle=0]{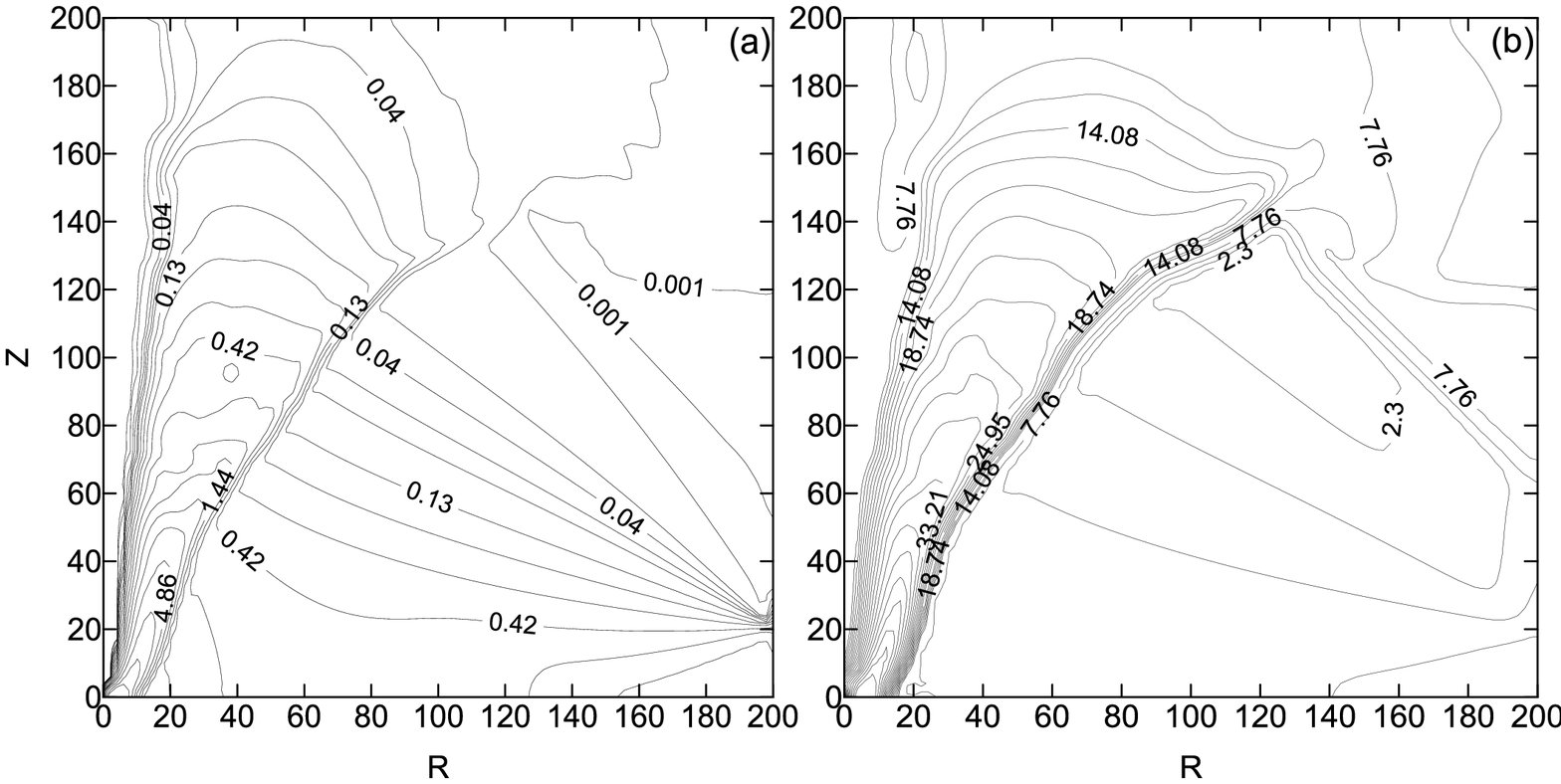}}
\caption{Density (a) and temperature (b) contours inside the halo ($\lambda=1$) in the absence 
of Compton cooling. Densities are in normalized unit and temperatures are in keV. See, text for details.}
\label{fig7ab}
\end{figure}

\begin{figure}
\centering{
\includegraphics[height=12truecm,width=12truecm,angle=0]{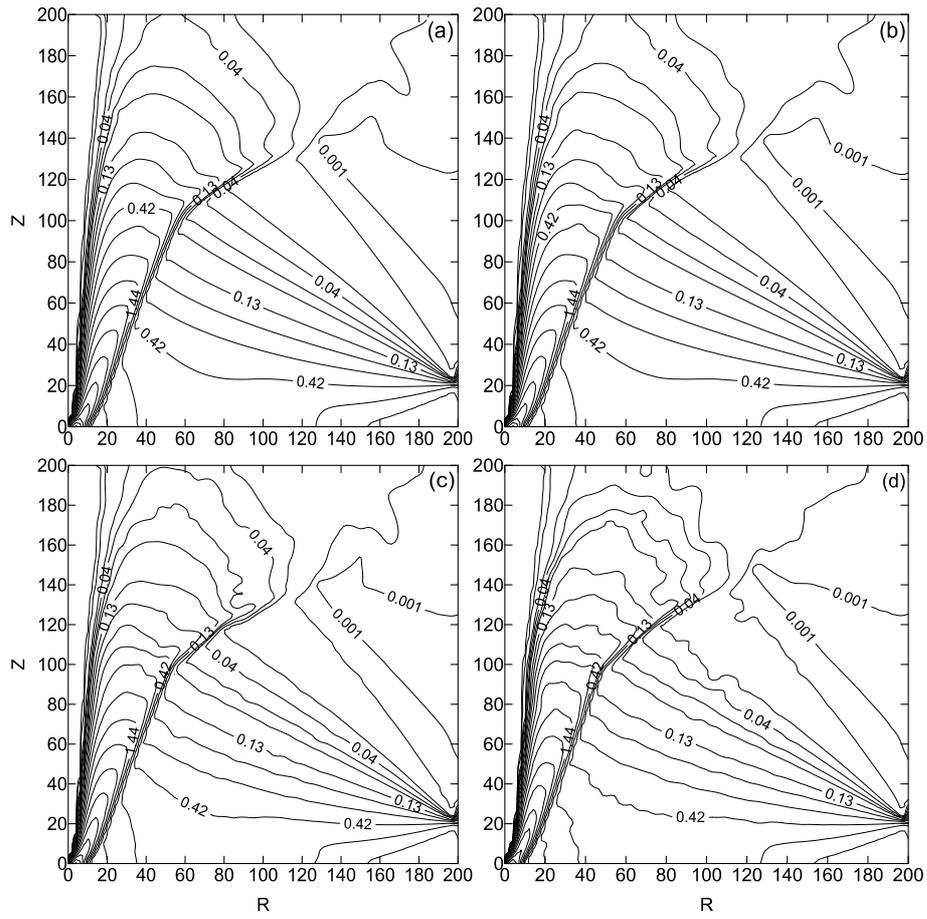}}
\caption{Change in the density contours in presence of cooling ($\lambda=1$). 
See, text for details. The conical region between the axis and shock wave becomes denser
as the accretion rate of the Keplerian disk is increased.}
\label{fig8abcd}
\end{figure}

We now turn our attention to the case where the cloud is formed by a low angular momentum flow. In this 
case the flow is already axisymmetric and due to centrifugal force a weak shock wave, or at least 
a density wave would be formed. In Fig. \ref{fig7ab}(a-b), we show the contours of constant density 
(Fig. \ref{fig7ab}a) and temperature (Fig. \ref{fig7ab}b) when no radiative transfer is included. 
Here the specific angular momentum of $\lambda=1$ was chosen. Density contour levels are drawn from $0.001-55.35$ 
(the successive level ratio is $1.5$), $55.35-73.73$ (successive level ratio is $1.1$). 
Temperature contour levels are drawn from $2.3-11.64$ (successive level ratio 
is $1.5$), $11.64-64.71$ (successive level ratio is  $1.1$). 
We note that a shock has been formed which bends outwards 
away from the equatorial plane (Ryu, Chakrabarti \& Molteni, 1997; Giri et al., 2010.).
In Fig. \ref{fig8abcd}(a-d), we show results of inserting a Keplerian disk    
in the equatorial plane. The inner edge is located at 3 $r_g$, the marginally 
stable orbit. Here, $\dot{m}_h = 1$ and $\dot{m}_d = $(a)  $1 $, (b) $ 2 $, (c)  $5 $ and (d)  $10 $ 
respectively (Cases 2(a-d) of Table 1). The densities used to draw the contours are the same as that in 
Fig. \ref{fig7ab}a. As the Keplerian disk rate is increased, the intensity of the soft photons interacting with 
the high optical depth (post-shock) region is increased. In Fig. \ref{fig8abcd}d, 
we observe that the conical region around the axis is      
considerably cooler. Thus, the density around the shock is enhanced. However, 
most importantly, with the increase in disk accretion rate, i.e., cooling, the shock location 
moves in closer to the black hole. This result has been already shown in the  
context of the bremsstrahlung cooling (Molteni et al. 1996).
\begin{figure}
\centering{
\includegraphics[height=12truecm,width=12truecm,angle=0]{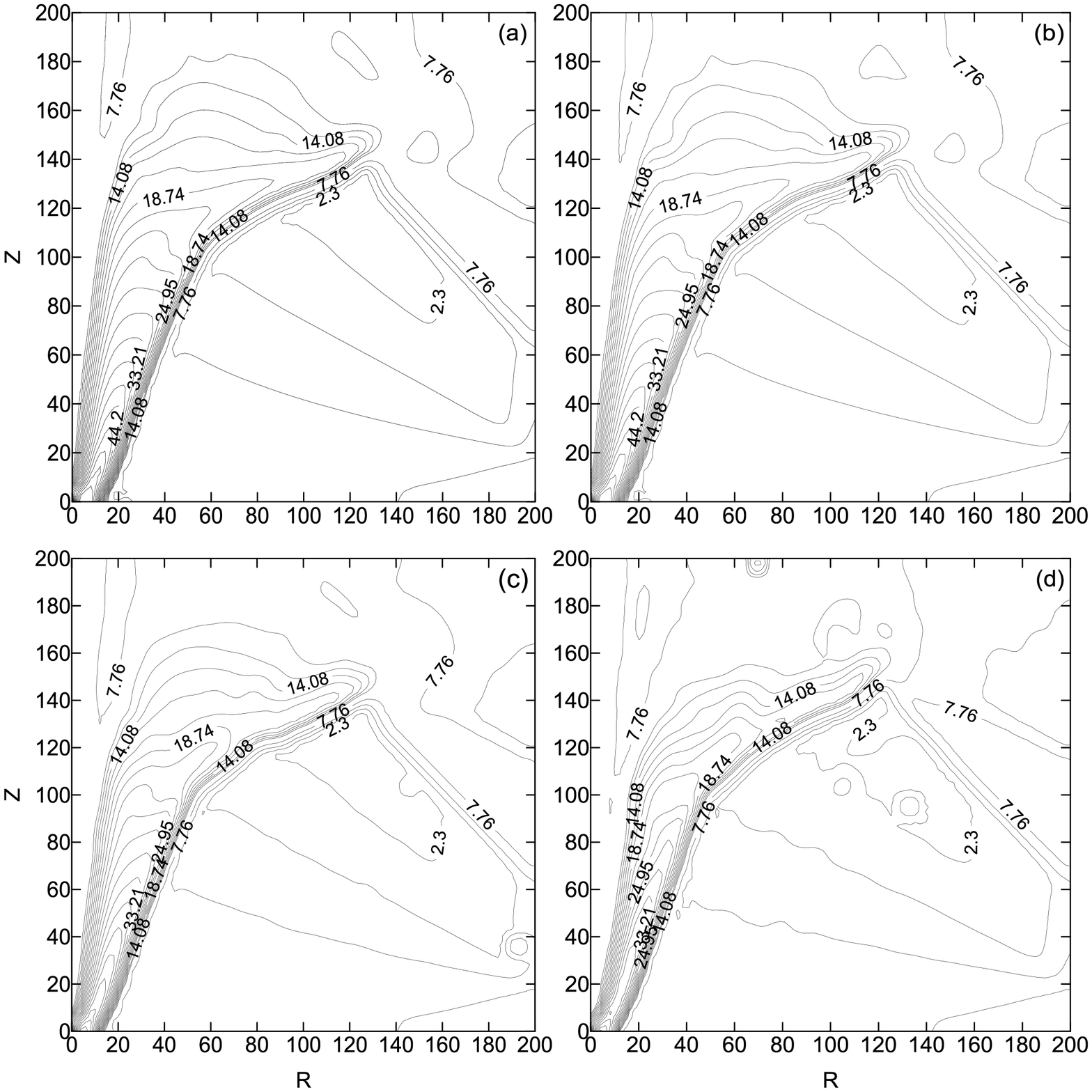}}
\caption{Change in the temperature contours in presence of cooling. The parameters 
are the same as in Fig. \ref{fig8abcd}(a-d). The temperature values used to draw the contours are 
the same as in Fig. \ref{fig7ab}b. Note that the shock shifts closer to the axis with the increase in
disk accretion rate.}
\label{fig9abcd}
\end{figure}
In Fig. \ref{fig9abcd}(a-d), we present the corresponding 
temperatures. The parameters are the same as in Fig. \ref{fig8abcd}(a-d) and the temperatures 
used to draw the contours are the same as that in Fig. \ref{fig7ab}b. 
The Comptonization in the shocked region cools it down considerably. Otherwise, 
not enough visible changes in the thermodynamic variables are seen. 
To understand the detailed effects of the radiative transfer on the 
dynamics of the flow, we take the differences in the pressure and velocity at each  grid
point of the flow for Cases 1d and 2d of Table 1. 
\begin{figure}
\centering{
\includegraphics[height=6truecm,width=12truecm,angle=0]{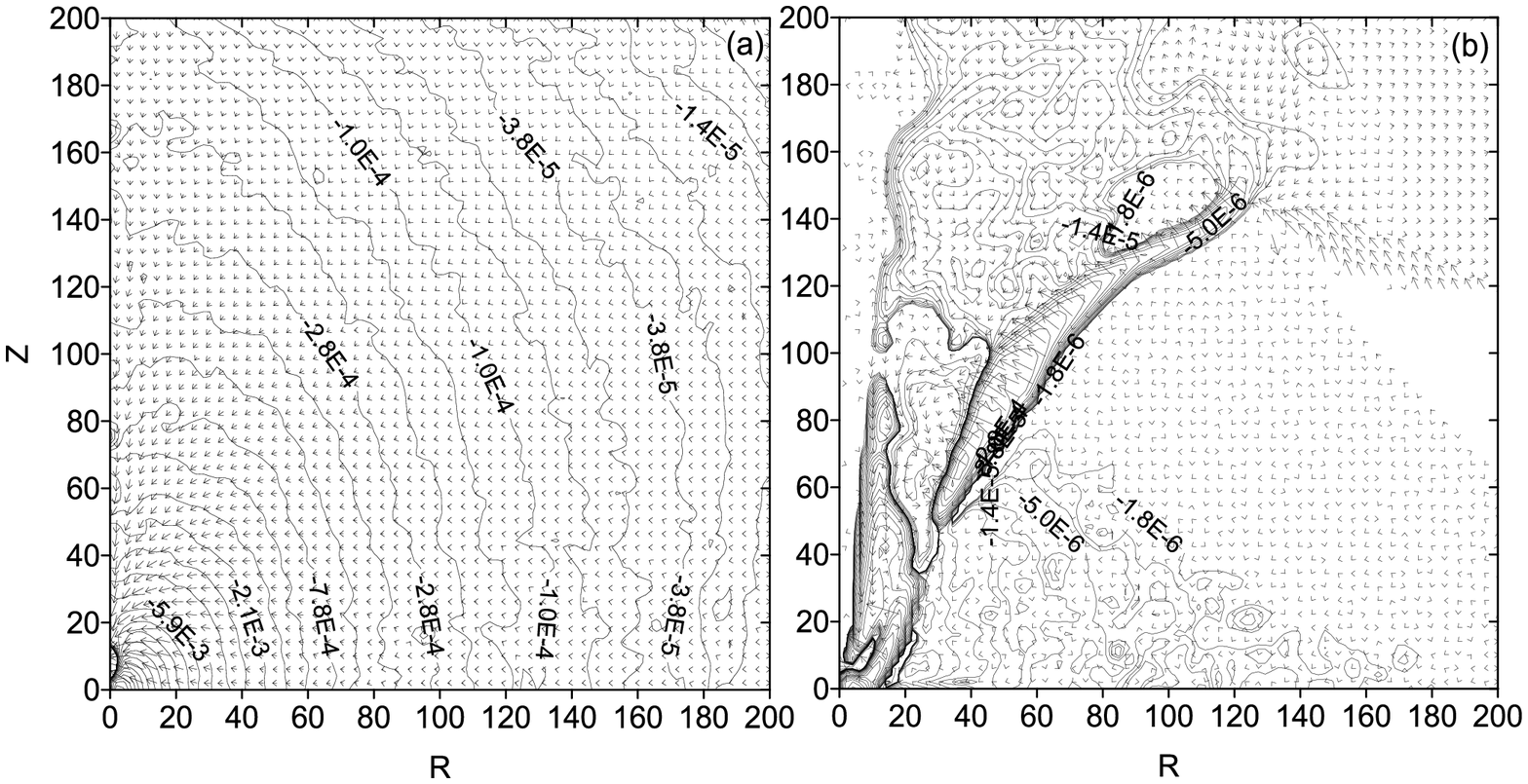}}
\caption{Difference in pressure and velocities between the flow with Comptonization and 
without Comptonization. Other parameters remain exactly the same. The cases are 
(a) Case 1d and (b) Case 2d of Table 1 respectively.} 
\label{fig10ab}
\end{figure}
In Fig. \ref{fig10ab}(a-b) we show the difference between the results of a purely hydrodynamical flow 
and the results by taking the Comptonization into account. Fig. \ref{fig10ab}a is for the flow with no 
angular momentum and Fig. \ref{fig10ab}b is drawn for the specific angular momentum $\lambda=1$. The 
contours are of constant $\Delta P= P_c-P_a$, where $P$ is the pressure and the subscripts $c$ and 
$a$ represent the pressure with and without cooling respectively. The arrows represent the difference
in velocity vectors in each grid. As expected, in both the cases, the changes are maximum near the axis. 
The fractional changes in pressures and velocities are anywhere between $\sim 0$ (outer edge) and $\sim 25$\% 
(inner edge and near the axis). Because of shifts of the shock location towards the axis, the 
variation of the velocity is also highest in the vicinity of the shock. Most importantly, the 
matter starts to fall back after losing the outward drive. This is the Chakrabarti-Manickam mechanism
(2000) which is believed to decide the nature of the light curves of objects containing 
outflows. Thus we prove that not only the symmetry is lost by the insertion of an axisymmetric 
soft photon source, the cooling process also plays a major role in deciding the dynamics of the flow.

\begin{figure}
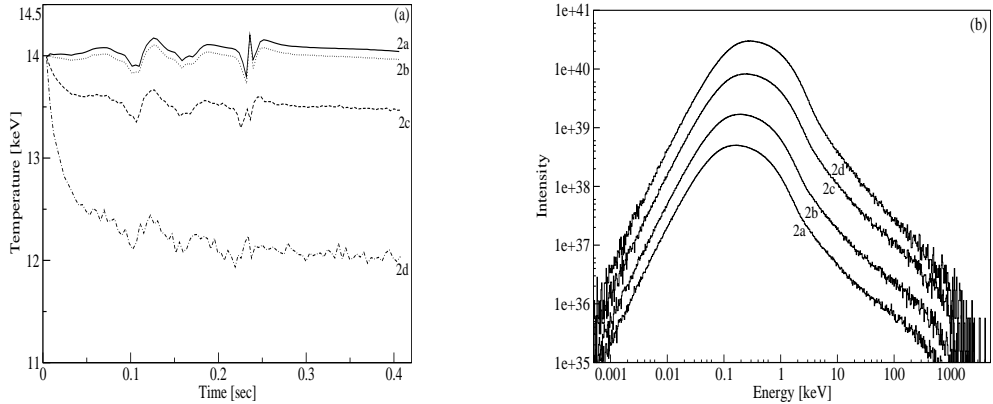

\centering{
\includegraphics[height=5.3truecm,width=5.3truecm,angle=0]{Fig11a.eps}
\hskip 1.5cm
\includegraphics[height=5.3truecm,width=6.0truecm,angle=0]{Fig11b.eps}}
\caption{Variation of (a) average temperature of the Compton cloud with iteration time and 
(b) spectrum with the increase of disk accretion rate. $\lambda = 1$ and 
$\dot{m}_h = 1$ are used. The solid, dotted, dashed and dash-dotted plots are for $\dot{m}_d = 1$, 
$2$, $5$ and $10$ respectively. With the increase in the disk rate, the temperature of the Compton cloud 
saturates at lower temperature. The solid, dotted, dashed and dash-dotted curves show the spectrum
for $\dot{m}_d = 1$, $2$, $5$ and $10$ respectively. The spectrum is softer for higher value of $\dot{m}_d$.}
\label{fig11ab}
\end{figure}

We now turn our attention to the dynamical variables and spectral behaviour of the rotating flow. 
In Fig. \ref{fig11ab}a, we show the variation of the average temperature of the 
Compton cloud as a function of the iteration time of the coupled code (Cases 2(a-d), Table 1). 
With the increase in the disk rate, the temperature of the Compton cloud saturates at 
a lower temperature. Fig. \ref{fig11ab}b shows the effect of the decrease in cloud temperature on the spectrum
due to the increase in disk rate. As we increase the disk rate, keeping the halo rate fixed, the spectrum 
becomes softer. 
\begin{figure}
\centering{
\vspace {1.0cm}
\includegraphics[height=5truecm,width=7truecm,angle=0]{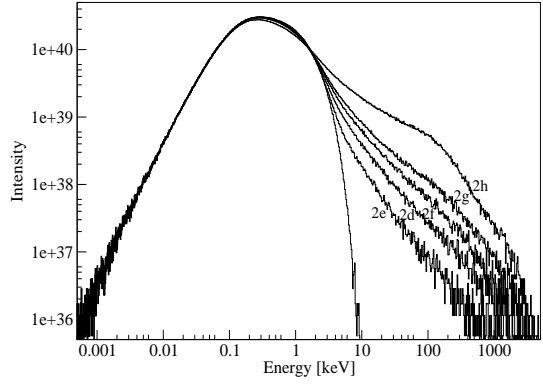}}
\caption{Variation of the spectrum with the increase of the halo accretion rate, keeping the 
disk rate ($\dot{m}_d = 10$) and angular momentum of the flow ($\lambda = 1$) fixed. 
The case number for which a spectrum is drawn is marked on it. The unmarked 
plot is the injected spectrum. The resulting spectrum becomes harder for the higher values of $\dot{m}_h$.}
\label{fig12}
\end{figure}

In Fig. \ref{fig12}, we show the effects of the increase of the electron number 
density (due to the increase of $\dot{m}_h$) for a fixed disk rate. The spectrum becomes harder as we 
increase the halo rate keeping the number of injected soft the photons the same.

\begin{figure}
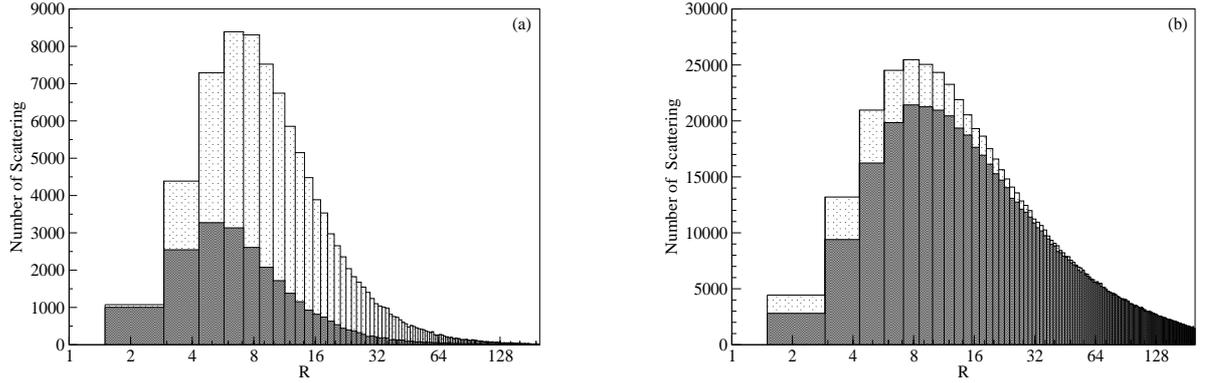

\vspace {1.0cm}
\centering{
\includegraphics[height=5truecm,width=7truecm,angle=0]{Fig13a.eps}
\hskip 1.5cm
\includegraphics[height=5truecm,width=7truecm,angle=0]{Fig13b.eps}}
\caption{Number of scatterings inside the spherical shell between $R$ to $R + \delta R$ ($ \delta R \sim 1.4$). 
The light and dark shaded histograms are for the cloud with and without bulk velocity, respectively. 
(a) Only the photons emerging from the cloud with energies $E$, where $50$ keV $< E <150$ keV, 
are considered here. (b) All the photons emerging from the cloud are considered here. Parameters used: 
$\dot{m}_d=1$,  $\dot{m}_h=10$ and $\lambda=0$.}
\label{fig13ab}
\end{figure}
We observe that the emerging spectrum has a bump, especially at higher accretion rates of the halo, 
at around $100$ keV (e.g. the spectra marked 1g, 1h in Fig. \ref{fig6abcd}a and the 
spectrum marked 2h in Fig. \ref{fig12}). A detailed 
analysis of the emerging photons having energies between 50 to 150 keV was made to see 
where in the Compton cloud these photons were produced. In Fig. \ref{fig13ab}a, we present the number of scatterings 
inside different spherical shells within the electron cloud suffered by these 
photons ($50 < E < 150$ keV) before leaving the cloud. Parameters used: 
$\dot{m}_d=1$,  $\dot{m}_h=10$ and $\lambda=0$. The light and dark shaded histograms are for the cloud with 
and without bulk velocity components, respectively. We find that the presence of bulk motion 
of the infalling electrons pushes the photons towards the hotter and denser 
[Figs. \ref{fig2ab}(a-b)] inner region of the cloud to suffer more and more scatterings. 
We find that the photons responsible for the bump suffered maximum number of scatterings around 
8 $r_g$. From the temperature contours, we find that the cloud 
temperature around 8 $r_g$ is $\sim 100$ keV. This explains the existence of the bump.
In Fig. \ref{fig13ab}b, we consider all the outgoing photons independent of their energies. 
The difference between the two cases is not so visible. This shows that the bulk velocity
contributes significantly to produce the highest energy photons. 
\begin{figure}
\vspace {1.0cm}
\centering{
\includegraphics[height=5truecm,width=7truecm,angle=0]{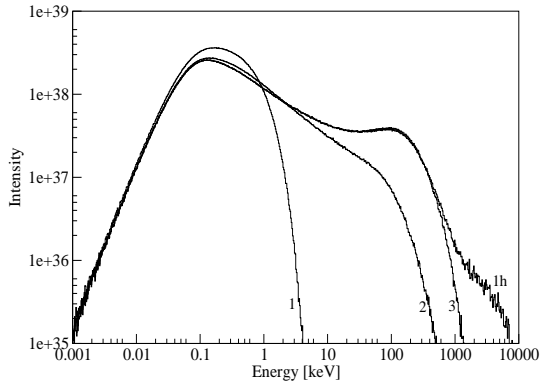}}
\caption{The spectrum for the Case 1h. The curves marked 2 and 3 give the spectra when the 
bulk velocity of the electron is absent for the whole cloud and for the cloud inside 3 $r_g$, respectively.
The curve marked 1 gives the injected spectrum. The bulk motion 
Comptonization of the photons inside the 3 $r_g$ radius creates the hard tail. The 
bump near 100 keV is a combined effect of the temperature and bulk velocity of the rest of the cloud.}
\label{fig14}
\end{figure}

In Fig. \ref{fig14}, we explicitly showed the effects of the bulk velocity on the spectrum. 
We note that the bump disappears when the bulk velocity of the electron cloud 
is chosen to be zero (Curve marked 2). This fact shows that the region around 8 $r_g$ in
presence of the bulk motion behaves more like 
a black body emitter, which creates the bump in the spectrum. Since the photons are 
suffering large number of scatterings near this region (8 $r_g$), most of them
emerge from the cloud with the characteristic temperature of the region. The 
effect of bulk velocity in this region is to force the photons to suffer larger 
number of scatterings. This bump vanishes for lower density cloud (low $\dot{m}_h$) 
as the photons suffer lesser number of scatterings. 
The photons which are scattered close to the black hole horizon and escape
without any further scattering, produce the high energy tail in the output spectrum. 
Curve 3 of Fig. \ref{fig14} shows the intensity spectrum of Case 1h (Table 1), 
when there are zero bulk velocity inside 3 $r_g$. We find that in the 
absence of bulk velocity inside 3 $r_g$, the high 
energy tail in the Curve 1h vanishes. This is thus a clear signature of the presence of bulk 
motion Comptonization near the black hole horizon.

\section{Summary and Discussions}

In this paper, we have extended our previous work using Monte Carlo simulations (Ghosh et al. 2009,
2010) to include the effects of Comptonization on the dynamics of the 
accreting halo having zero and very low angular momentum. 
We studied the properties of the emerging spectrum from the Chakrabarti-Titarchuk
model of a two component flow, one component being the 
Keplerian disk on the equatorial plane and the other component is the low angular momentum accreting halo, 
which is acting as the Compton cloud.
In Table 1, we have given the parameters of all the cases which were run. We
note that as we enhance $\dot{m}_d$, $N_{inj}$ ($\sim \dot{m}_d^{3/4}$; see, \S 2.2) 
is also enhanced, increasing the number of photons $N_{sc}$ undergoing Compton scattering. 
If we keep the halo rate $\dot{m}_h$ fixed, then increasing $\dot{m}_d$ increases $N_{unsc}$, 
the number of photons escaping from the disk while keeping $p$ almost unchanged. 
For Cases 1(a-d) and 2(a-d), the percentage of photons undergoing scattering 
$p$ is $\sim 20 \%$ and $\sim 19 \%$, respectively. When we increase $\dot{m}_h$, 
keeping $\dot{m}_d$ constant, this percentage increases rapidly due to the decrease in $N_{unsc}$. We see 
from Table 1, as $\dot{m}_h$ is increased from $0.5$ to $10$, keeping $\dot{m}_d = 1$ 
[Cases 1e, 1a, 1(f-h)] percentage of scattered photon increases from $\sim 11 \%$ to 
$\sim 76 \%$. The same situation prevails for the $\lambda = 1$ cases 
[Cases 2e, 2d, 2(f-h)], where $p$ increases from  $\sim 11 \%$ to $\sim 51 \%$, for the increase in 
$\dot{m}_h$ from $0.5$ to $5$, keeping $\dot{m}_d = 10$. $N_{bh}$ remains almost constant if we 
keep the halo rate constant. If we increase the halo rate, $N_{bh}$ increases rapidly, because 
increase in $\dot{m}_h$ increases the density of the cloud and thus pushes the photons
towards the black hole. As we increase $\dot{m}_d$, the cooling time $t_0$ decreases, since
with the increase of $\dot{m}_d$ number of soft photon increases. Thus, the cloud cools 
down at a faster rate. We also notice that $t_0$ decreases as $\dot{m}_h$ increases, due to
the increase of $N_{sc}$. Spectral index $\alpha$ increases with the disk rate for
a fixed halo rate, and it decreases with halo rate for a fixed $\dot{m}_d$. This can be explained 
by the fact that as we increase $\dot{m}_d$, the electron cloud becomes cooler, the spectrum gets softened. 
On the other hand, when we increase the number of hot electron inside the cloud (i.e. $\dot{m}_h$), 
for the same $N_{inj}$ we get a hotter system. This makes the spectrum harder. These results
are consistent with the Chakrabarti-Titarchuk scenario of two component accretion.

Our major conclusions are the followings: 

\noindent i) In the presence of an axisymmetric disk which supplies soft photons to the 
Compton cloud, even an originally spherically symmetric accreting Compton cloud 
becomes axisymmetric. This is because, due to the higher optical depth, there is a significant
cooling near the axis of the intervening accreting halo between the disk and the axis. 

\noindent ii) Due to the cooling effects close to the axis, the pressure drops significantly, which 
may change the flow velocity up to $25$\%. This effect becomes more for
low angular momentum flows which produces shock waves close to the axis. The post-shock region
cools down and the outflow falls back to the disk. This shows that the Chakrabarti-Manickam
(2000) mechanism of the effects of Comptonization on outflows does take place.

\noindent iii) The emitted spectrum is direction dependent. The spectrum along the axis shows a
large soft bump, while the spectrum along the equatorial plane is harder. 

\noindent iv) Photons which spend more time (up to 100 ms in the case considered) inside 
the Compton cloud produce harder spectrum as they scatter more. However,
if they spend too much (above 100 ms) time, they transfer their energies back to 
the cooler electrons while escaping. These results would be valuable for interpreting 
the timing properties of the radiation from black hole candidates.

\section{Acknowledgment}

HG thanks ISRO RESPOND project for support. 
The authors thank P. Laurent and D. Ryu for providing us with basic Monte Carlo and hydrodynamics codes
and providing helpful comments.

{}
\end{document}